\documentclass[submission]{eptcs}
\providecommand{\event}{CAPNS 2018} 
\usepackage{underscore}           

\usepackage{graphicx}

\usepackage{physics}
\usepackage{longtable} 
\usepackage{setspace}

\title{Quantum Semantic Correlations\\ in Hate and Non-Hate Speeches\thanks{This paper has been made possible thanks to a corpus of hate and non-hate speeches shared by the Berkeley D-Lab.}}
\author{Francesco Galofaro
\institute{Politecnico\\Milano, Italy}
\institute{Free University of Bozen\\ 39100 Bozen, Italy}
\email{francesco.galofaro@polimi.it}
\and
Zeno Toffano\qquad\qquad Bich-Li\^{e}n Doan
\institute{CentraleSup\'{e}lec\\ Gif-sur-Yvette, France}
\email{\quad zeno.toffano@centralesupelec.fr \quad\qquad Bich-lien.doan@centralesupelec.fr }
}
\def\titlerunning{Quantum Semantic Correlations}
\def\authorrunning{F. Galofaro, Z. Toffano \& B.-L. Doan}
\begin{document}
\maketitle             

\begin{abstract}
This paper aims to apply the notions of quantum geometry and correlation to the 
typification of semantic relations between couples of keywords in different documents. In particular
we analysed texts classified as hate / non hate speeches,  containing the keywords \textit{women}, \textit{white}, and \textit{black}. 
The paper compares this approach to cosine similarity a classical methodology, to cast light on
the notion of ``similar meaning''. 
\end{abstract}

\section{Corpus}
The Online Hate Index\cite{OHI}, a joint initiative 
of Anti Defamation League\textsc{\char13}s Center for Technology and Society
 and UC Berkeley\textsc{\char13}s D-Lab, consists
of 7619 comments from the platform Reddit, collected in 2016 during the USA Presidential campaign. While the corpus has been funded by ADL, no research activities are conducted by ADL or directed based on their personal goals: research design and testing has been conducted entirely by Berkeley D-Lab, an interdisciplinary academic unit. The comments were labelled by a team of undergraduates under the supervision and using methods developed by D-Lab. ADL had no involvement in the selection of comments, development of methods, and labels applied to comments.
7184 comments have been manually labelled ``non-hate'', whereas 411 have been considered as ``hate'' speeches
with the purpose of stimulating further machine learning researches. The five most used words in the collected hate speeches
are \textit{Jews}, \textit{white}, \textit{hate}, \textit{black}, and \textit{women}. Hate speeches present also some peculiar features:
the average number of words, the average number of all caps, and the average sentence length 
is higher. Among the five most used words in hate speeches, \textit{white} and \textit{black} are interesting because they can be
considered an antonymic couple from the point of view of lexical semantics. For this reason, we decided to analyse the semantic relations between the terms
\textit{white}, \textit{black}, and \textit{women}.
\subsection{Sub-corpora}
As the corpus is subdivided into hate (see table ~\ref{tab1}) and non-hate speeches (see table ~\ref{tab2}), we identified six sub-corpora:
\begin{itemize}
	\item H.WBWh: hate speeches corresponding to the logical expression $black*white*women$;
	\item H.WhW-B: $white*women-black$ hate speeches; 
	\item H.BW-Wh: $black*women-white$ hate speeches; 
	
	\item NH.WBWh: $black*white*women$ non-hate speeches;
	\item NH.WhW-B: $white*women-black$ non-hate speeches; 
	\item NH.BW-Wh: $black*women-white$ non-hate speeches; 
	
\end{itemize}

As we can see, selected hate-speeches tend to use the words \textit{white}, \textit{black}, and \textit{women} in the same document, whereas not-hate speeches tend to use only \textit{white},  \textit{women} or \textit{black},  \textit{women}. It is sometimes difficult to understand the criteria that have been used for labelling the corpus. In particular, NH.WBWh.2 and NH.WhW-B.4, two documents labelled as non-hate, though avoiding offensive terms contain stereotypes on the relation between race and IQ. We hope that a semantic analysis will be helpful to re-discuss the classification in case of mistakes, e.g. if a document classified as non-hate speech shows the same semantic features of hate speeches.

\subsection{Note on the formalism}
The logical formalism we used to represent elementary semantic relations corresponds both to Boole\textsc{\char13}s algebra and to the one proposed by Greimas and Rastier in their well-known semiotic square\cite{square} and suggests a strong connection to Dubois and Toffano\textsc{\char13}s eigenlogic which could provide a formal description of the semantic relations immanent  to the text, thus allowing the formulation of semantic queries\cite{eigen}. 

\begingroup
\scriptsize 
\begin{longtable}{|p{15.7cm}|}
\caption{Hate speeches. Italic underlines the occurences of the keywords.}
\label{tab1}\\
\hline
Query: $black*white*women$\\
\hline
\textbf{H.WBWh.1 -} Did you ever notice how these anti-\textit{white} racist \textit{black}s always pick on the old, the weak or the infirm? Old ladies. Children. Young \textit{women}. Teenage geeks. And they almost never do it alone, but run in packs, like jackals or hyenas.\\
\textbf{H.WBWh.2 -} Eastern Europe will likely be the most reliable backup plan for most \textit{white}s. And while most of western Europe is cucked out right now, I have higher long term hopes for there than America. Most Americans are satisfied with only going so far as civic nationalism. Nationalism is intertwined with ethnicity in Europe and so ethnonationalism is the inevitable result from the resurgence of the right there. I expect that America's nationalism will never again become attached to race, unfortunately. The elites in all western countries are banking on a multiracial future and in many cases are mating accordingly. This girl has mostly \textit{white} features and in all likelihood is ~75\% \textit{white} (given that most American \textit{black}s are 20-40\% \textit{white} themselves), but the media will play up her \textit{black}ness in order to make it seem that \textit{black} \textit{women} are desirable.\\
\textbf{H.WBWh.3 -} It was always a lot like this.  There were always elements of it.  For example it was immidiately anti-male from the word go.  It was just easier to campaign that way because society was so anti-male / pro-female already.  Hence the saying that feminism is just the other side of the coin from traditional conservatism.
For example feminists went on and on about drunk husbands beating their poor wives the way they go on about rape today.  A good false accusation makes for fine propaganda and feeds into what everyone likes to think about men anyway.
Feminists could have taken the high road.  They didn't.  They took the easy screw men over, sex war men and \textit{women} are enemies and men are all evil road.
So the rot was there from the beginning.  So was the narcissism.  The victimry.  The ``we rich \textit{white} \textit{women} are worse off than \textit{black} slaves'' crap.  Reality was they were treated better than anyone else.  but that doesn't sell well.  Victimry does (for \textit{women}).  They shamelessly used their power to get more.
Sure some wanted equal responsibilities and duties too at first but those voices, never a majority, got fewer and fewer.  Getting more privileges was just the easy path.
So yes to the vote, but no to conscription that went with it.
And you can see how close to a hate movement all that already was even back in the day when they had some issues that were actually worth fixing (even if they were not a matter of equality because men's issues were not being addressed too).
But since they got all their issues dealt with by around 1890 (except the vote because most \textit{women} still opposed the vote for \textit{women} until the 1910s) they became a movement without a cause and just wnet off the deep end of anti-male hate.\\
\textbf{H.WBWh.4 -} Sometimes I feel like those movements became obsolete the moment \textit{women} got equal rights with men and people stopped thinking about \textit{black}s as of inferior race. Now they just keep momentum, turning \textit{women} and minorities into privileged classes. If they keep this up in a few decades we would *need* MRA and \textit{white} rights activists.\\
\hline
Query:  $white*women-black$\\
\hline
\textbf{H.WhW-B.1 - } Those 20 \textit{women} ought to be quarantined in a special zoo and denied treatment for their HIV. Then every \textit{white} woman should be forced to walk through that zoo to see those \textit{women} slowly die from race-treason. These whorish \textit{women} need to be brought back into line, they will be the death of our race. \\
\textbf{H.WhW-B.2 - } Jewish gangs kidnapping/raping/killing \textit{white} children as young as 2 years old to sell snuff films](http://www.jewwatch.com/jew-atrocities-blood-libel.html)  , then Jews had the Italian news executives fired for AIRING THE STORY (Jews HATE being called out on crime and DEMAND FULL CONCEALMENT or else it's ``ANTI SEMITIC!!!''....Jews think it's ANTI SEMITIC to PROTECT YOUR CHILDREN FROM THEM).
[Israel's \$1 billion a year sex slave trade of \textit{white} \textit{women}](http://www.jewlicious.com/2006/04/because-we-were-slaves-israels-sex-trade/) \\
\textbf{H.WhW-B.3} There is no need to side with kikes against \textit{white} \textit{women} who put their money where their mouth is and goes and protests the jew at her own risk. 
Why on earth do I care if the Jews are weakening someone who is an avowed enemy, like Hamas?  Granted they're not capable of it, but they've made declarations of intent for conquering Italy and other places in Europe.  The Muslims will forever wage their jihad against Europe regardless of whether or not Israel exists.  Their religion is one of rape and plunder.
I don't really protest Israel because I couldn't care less what Israel does in Israel.  If the time comes for Europe to reclaim Turkey and the Levant, then I will care.  
I care what Israeli stooges do here in the west, what propaganda they sow, what legislation they lobby for, etc.\\
\hline
Query:  $black*women-white$\\
\hline
\textbf{H.BW-Wh.1 -} Based on the many, many videos I've watched of chimpouts, \textit{black} \textit{women} are more aggressive and more violent than \textit{black} men. They seem to think there are no consequences for them when they punch other people in the face.\\
\textbf{H.BW-Wh.2 -} Liberals only teach the bad in american history. I had multiple teachers that told me that slavery affects \textit{black} people today and \textit{women} only make 70 cents to a man. These are both lies, and there is nothing taught about how we spread ideas of individual freedom across the western world and gave more rights to \textit{women}, minorities,  plants and animals than any other, all thanks to ``racist slave holders'' so yeah, teach slavery all you want, but also include the fact that these ideas were not constitutional and mostly pushed by democrats.\\
\hline
\end{longtable}
\endgroup

\begingroup
\scriptsize
\begin{longtable}{|p{15.7cm}|}
\caption{Non-hate speeches. Italic underlines the occurrences of the keywords.}
\label{tab2}\\
\hline
Query: $black*white*women$ \\
\hline
\textbf{NH.WBWh.1 -} I think mansplaining might be a real thing.
Explaining done by a man is a real thing. But why make it gendered? \textit{women} explain things to men all the time, at times also using a patronizing tone. Anecdotally, I've experience more \textit{women} explaining something in a patronizing way than I have men. 
Even if there would be statistics that show men do it more often in a patronizing way (which there aren't), it's still hard to argue for making it gendered. 
To put it in an analogy: we know that \textit{black} people in the USA commit more burglary than \textit{white} people (in relative terms). Should we call it ``\textit{black}burgling''? I don't think so: I think it implies that comitting burglary is somewhat characteristic of \textit{black} people, which isn't the case since it's only a minority of \textit{black} people doing it. It would probably also help drive the wedge even further between \textit{black} and \textit{white} people. And what would have been gained? A fancy new word to insult \textit{black} people with? Why can't it just be ``burglary (done by a \textit{black} person)''? The same goes here.\\
\textbf{NH.WBWh.2 -} Well, you're not wrong. \textit{black}s, men and \textit{women}, are of significantly lower intelligence than all other races on the planet, with the single exception of the aborigines of Australia, who are just as limited, mentally, as \textit{black} Africans.
I'm talking about a BIG difference in intelligence, not a small difference. The average IQ of \textit{white}s is around 100. Hispanics come in at around 89. American \textit{black}s, who have mingled their genetics with \textit{white}s for generations of interbreeding, come in at an average of 85. African \textit{black}s have an average IQ of 70! That's right, 70. By normal \textit{white} standards, the average negro in Africa is mentally retarded.
These IQ tests have been done many different times, many different ways, all around the globe, and they all show the same thing. They are not  wrong.
Since  you are \textit{black} yourself, I should point out that a general or average IQ for a race has no bearing on the IQ of any individual in that race. By that I mean you yourself may be a genius. There are many \textit{black} geniuses. I am talking here about average intelligence.
Why is average intelligence important? Because it dictates what a race, as a whole, is able to accomplish. If you look at Africa, \textit{black}s were able to build almost nothing. No roads. No wagons to drive on those non-roads. They didn't even learn to domesticate horses to ride by themselves. Everything African negroes have, they learned from other races. And please don't talk about Egypt. The peoples of northern Africa were not negroes in ancient times, and to a large extent are still not negroes. They were Mediterranian peoples, like the Greeks, Etruscans and Romans.
Why is the lack of accomplishment of African \textit{black}s in history important today? Because \textit{black}s still cannot create or build anything of importance, as a race. Their low intelligence, coupled with other negative factors that have been less well-demonstrated, prevent them from achieving anything.
 Just look at any \textit{black} city in America. Any \textit{black} city, take your  pick. As soon as a city goes \textit{black}, it goes to ruin and decay. Always.\\
\hline
Query: $white*women-black$ \\
\hline
\textbf{NH.WhW-B.1 -}  8chan trolls are using it to abuse \textit{women}, children, minorities, and gays Hetero \textit{white} men can't be abused, folks. If the very same thing happens to them it's... well... irrelevant.\\
\textbf{NH.WhW-B.2 -} Many \textit{white} \textit{women} voted for Trump not because they were concerned about their income, but because they are the ?Ivanka voters?. Their vote was for Ivanka instead of Donald. They love her style and success story.
I can't imagine this being true at all.\\
\textbf{NH.WhW-B.3 -} Then there's \textit{white} ribbon day. With posters saying 'Are you man enough to stop violence against \textit{women}?'
http://\textit{white}ribbon.org.nz/\\
\textbf{NH.WhW-B.4 -} It's become a sort of cliche that Hispanic \textit{women} apparently urge their daughters to seek out \textit{white} men (or at the very least, lighter skinned men) in order to ``improve the race''. There is about a 10 point IQ difference between \textit{white}s and most mestizos, and fairer skin is usually seen as more attractive regardless of political beliefs, so their efforts aren't misguided.
The last time I was in Corpus Christi, as well as when I visited Phoenix and San Diego, it seemed like every single young \textit{white} male had a mestizo girlfriend. And in their minds ``why not?'' An average \textit{white} guy can get a high end mestizo girl easier than an average \textit{white} girl, and chances are that she is eager to please because she is happy just to have a \textit{white} guy.
The rate of WM/HF mixing in the southwest is every bit as bad as WM/AF mixing on the West Coast. Here in the Midwest the most commonly seen mixed couple is BM/WF so I was rather shocked when I've gone out west and seen the extent of what is happening out there.\\
\textbf{NH.WhW-B.5 -} How would energising the base help. He already has them. ``You would be in jail'' ``hate in your heart'' and the ``devil'' comment are all for the base and he still dropped in the polls afterwards. 
The guy needs greater support from \textit{women}, he needs greater support from minorities. As far as I know registration hasnt surged for the non college educated \textit{white} voters so he needs that as well.
Currently clinton can run out the clock, she won't and isn't trying that but she could rather comfortably not do much and let trump focus on attacking his accusers.
I have a feeling he is going to leave that arguement behind now given that he has called the accusers horrible horrible liars and 
?Take a look. Look at her. Look at her words. And you tell me what you think. I don?t think so,?\\
\textbf{NH.WhW-B.6 -} I wonder what she would say to me. I'm a \textit{white} male that voted for Jill Stein/Ajamu Baraka. I voted for 2 \textit{women}, rather than her 1.\\
\hline
Query:  $black*women-white$\\
\hline
\textbf{NH.BW-Wh.1 -}  Win or lose, a lot of people have got the red pill. Trump just did not play the demographics well. Alienating \textit{black}s, mexicans, muslims,  \textit{women}.  But many of them also know what is happening, but they consider trump worse
The establishment has suffered from a huge loss in credibility which they can never recover. As an example see the comments on cnn, abc, nyt facebook pages\\
\textbf{NH.BW-Wh.2 -} Unlike jubbergun, i didn't even notice the race of the attackers. I think one of the men standing around was \textit{black}, but that was the only time i noticed color. Jubbergun is either a troll or a PC moron, that's all i can say. 
As for my specifically using bonobos rather than common chimps or gorillas, that is because the bonobos are our closest relatives and also one of the few species of primate amongst which the females commit most of the violence. Hence, these \textit{women} have unleashed their inner bonobo.\\
\textbf{NH.BW-Wh.3 -} I'll admit, I'm really pro sanders but this upset me. He was pretty rude to the woman, I wish he let her talk more and had more of a conversation.
To me, most the arguments I get into, the one who rages first is often the one with at least the most to lose and at most the least willed debater. Not to mention it is prejudice.
Urban? Come on. If humans fight, if we're in a war, we use guns. Plain and simple. The \textit{black} men in this country and \textit{black} \textit{women} in particular have been SHUT OUT of so many opportunities in this country and MANIPULATED into the war they think they have to wage. Manipulated by terrible, evil motherfuckers in badges who go around thinking they're doing the lord's work.
And really, really fucking unfortunately, I can't help but feel that Sanders' view as the socialist democrat is so far right and so agreed upon by so many... it really does show how ingrained gun culture is in our country.\\
\textbf{NH.BW-Wh.4 -} claims quoting Dr. Martin Luther King, Jr., to \textit{black} \textit{women} is a violent and ?cisheteropatriarchy? act.
You seriously can't make this up.\\
\textbf{NH.BW-Wh.5 -} That's probably because 30 years ago they were not bashing \textit{black}s or \textit{women}. Well, \textit{women} only got bashed if they mouthed off.\\
\hline
\end{longtable}
\endgroup

\section{State-of-the-art}
A geometric approach to semantic space studies has been proposed first by Jean Petitot, in terms of catastrophe theory\cite{petitot2004morphogenesis}. As quantum geometry is concerned, it is used by scholars in Information Retrieval for the purpose of unifying
vector, logic, and statistical approaches\cite{van2004geometry,Melucci}. Among others, Bruza and Woods\cite{Bruza}, applied it to the semantic
representation of polysemic words as a superposition state. Barros, Toffano, Meguebli and Doan\cite{bell}
proposed to interpret 
the notion of entanglement as a measure of the strength of the semantic relation between
two query-words, both present in a certain document. 
To this purpose, using the Hyperspace Analogue to Language (HAL) method\cite{Lund1996}, the authors formalised the semantic space of a document as a
square matrix, as we will explain hereafter. Many quantum information retrieval scholars prefer this technique because it is Hermitian and it allows the implementation of a density matrix \cite{van2004geometry,Melucci}. Instead of measuring cosine similarity between two keywords, the work in \cite{bell}
makes use of the Gram-Schmidt orthogonalisation method to measure the degree of correlation between
the words, characterized by the violation of a CSHS inequality \cite{clauser1969proposed}. Pushing forward this idea, Galofaro, Toffano, and Doan\cite{doi:10.1108/K-05-2017-0187} proposed a theoretical paper in which 
observables are interpreted as semantic features. The Born rule is used to find the expectation values associated to the application of a specific observable
to two word-vectors in order to measure the degree of correlation/anticorrelation between them\cite{susskind2014quantum}. The present paper aims to test this method, and to compare it with the classical cosine similarity measure.

\section{Relevance to language processing}
\label{sec:relevance}
It is possible to ask how are we going to interpret the correlation value in
terms of linguistic features. According to Umberto Eco\cite{Eco}, the terms ``semantics'' has been used in five different acceptations:
\begin{enumerate}
	\item Lexicology: a study of meaning outside every context (dictionary);
	\item Structural Semantics: interested in semantic fields considered as systems; 
	\item Study of the relation between the meaning and the referent; 
	\item Truth-conditional logic;
	\item Textual semantics: a study of the peculiar meaning assumed by terms
	and words in their context; 
	
\end{enumerate}

Though the five levels are obviously related, the text and the context have the last word in defining the meaning of terms. For example, according to any thesaurus, \textit{black} and \textit{white} are antonyms (if black, then not white and vice versa). Having a look at our corpus, we find: ``most American \textit{blacks are} 20-40\% \textit{white}'' (H.WBWh.2), weakening the antonymy. The HAL method allows us to work on semantics in sense of 5 because we formalise the semantic relations between terms that constitute a given context. These become the characteristics of a semantic space.

\subsection{Commutation test and quantum correlation}
With measuring the degree of quantum correlation we are searching for a semantic equivalent to Hjelmslev\textsc{\char13} commutation test. Commutation of elements of the \textit{expression plan} aims to search for linguistic units. If we substitute ``black'' with ``Afro-American'' in \textit{blacks, men and women are of significantly lower intelligence than all other races on the planet}, we notice how meaning is unaltered, while if we substitute it with ``Afghan hound'', the meaning changes. We could even suggest that this is why the original sentence is actually racist: we speak about men as they were dogs. However, \textit{*Afghan hounds, men and women} is not correct in English because of structural reasons related to semantics in sense 2:  ``men'' and ``women'' carry a structural \textit{classeme} (an element of meaning) $(human \to - animal)$\cite{greimas1983structural} . What if we were able to commute \textit{meanings}, and not \textit{signifiers}? For example, what if we were able to change the meaning unit ``human'' in ``dog'' while preserving  ``male'' and ``female'' all along the sentence? This is what we mean by ``commutation test on the content plan''. As a result of the test, such an abstract machine as a computer could probably generate, on the expression plan, a sentence like \textit{Afghan hounds, sires and bitches}. The Born rule provides a tool to measure the expectation values for these commutations.

\section{Design}
In synthesis, we prepared the corpus
reducing each word to its stem; we then applied the HAL method to obtain two word vectors representing the keywords we are interested in, and a document vector; finally, we measured the the cosine similarity of the keywords and their (anti)correlation value.

\subsection{Cleaning the corpus}
Since we are interested in every kind of semantic information not manifested by morphology or syntax, we used the Python library \textit{nltk Lancaster} stem to reduce different tokens to the same type (e.g. black, blacks, blackness). The \textit{Lancaster} stemmer is more aggressive than the alternative \textit{nltk Porter} stemmer, which can distinguish between \textit{woman} and \textit{women}. Obviously, a stem is not necessarily identical to its morphological root: our purpose is only to reconstruct the immanent net of relations underlying the manifest words. For a similar reason, we used ntlk stopwords list to eliminate syncategorematic terms. We also used regex to eliminate all
not relevant signs such as punctuation\cite{Zinoviev}.

\subsection{The matrix}
We applied the HAL method to each document of each sub-corpus to formalise it. Given $k$ roots occurring in the document, we calculate a $k \times k$ matrix which represents the semantic space of the document. For example, table~\ref{tab3} shows the HAL matrix of the document: \\

NH.BW-Wh.5: \textit{That's probably because 30 years ago they were not bashing blacks or women. Well, women only got bashed if they mouthed off}\\
\begin{table}[ht]
\centering
\caption{HAL Matrix corresponding to document NH.BW-Wh.5 (window: 11)}\label{tab3}
\resizebox{1\textwidth}{!}{\begin{tabular}{c|cccccccccccccccccccc}
 & {\bfseries 30} & {\bfseries ago} & {\bfseries bash} & {\bfseries becaus} & {\bfseries black} & {\bfseries got} & {\bfseries if} & {\bfseries mouth} & {\bfseries not} & {\bfseries off} & {\bfseries onli} & {\bfseries or} & {\bfseries probabl} & {\bfseries s} & {\bfseries that} & {\bfseries they} & {\bfseries well} & {\bfseries were} & {\bfseries women} & {\bfseries year}\\
\hline
{\bfseries 30}&10&0&0&9&0&0&0&0&0&0&0&0&8&7&6&0&0&0&0&0\\
{\bfseries ago} & 8 & 10 & 0 & 7 & 0 & 0 & 0 & 0 & 0 & 0 & 0 & 0 & 6 & 5 & 4 & 0 & 0 & 0 & 0 & 9\\
{\bfseries bash}&4 & 6 & 22 & 3 & 3 & 9 & 0 & 0 & 10 & 0 & 8 & 4 & 2 & 1 & 0 & 7 & 6 & 8 & 12 & 5\\
{\bfseries becaus} & 0 & 0 & 0 & 10 & 0 & 0 & 0 & 0 & 0 & 0 & 0 & 0 & 9 & 8 & 7 & 0 & 0 & 0 & 0 & 0\\
{\bfseries black} & 3 & 5 & 9 & 2 & 10 & 0 & 0 & 0 & 8 & 0 & 0 & 0 & 1 & 0 & 0 & 6 & 0 & 7 & 0 & 4\\
{\bfseries got} & 0 & 0 & 3 & 0 & 4 & 10 & 0 & 0 & 2 & 0 & 9 & 5 & 0 & 0 & 0 & 0 & 7 & 1 & 14 & 0\\
{\bfseries if} & 0 & 0 & 10 & 0 & 2 & 8 & 10 & 0 & 0 & 0 & 7 & 3 & 0 & 0 & 0 & 0 & 5 & 0 & 10 & 0\\
{\bfseries mouth} & 0 & 0 & 7 & 0 & 0 & 6 & 8 & 10 & 0 & 0 & 5 & 1 & 0 & 0 & 0 & 9 & 3 & 0 & 6 & 0\\
{\bfseries not} & 5 & 7 & 0 & 4 & 0 & 0 & 0 & 0 & 10 & 0 & 0 & 0 & 3 & 2 & 1 & 8 & 0 & 9 & 0 & 6\\
{\bfseries off} & 0 & 0 & 6 & 0 & 0 & 5 & 7 & 9 & 0 & 10 & 4 & 0 & 0 & 0 & 0 & 8 & 2 & 0 & 4 & 0\\
{\bfseries onli} & 0 & 0 & 4 & 0 & 5 & 0 & 0 & 0 & 3 & 0 & 10 & 6 & 0 & 0 & 0 & 1 & 8 & 2 & 16 & 0\\
{\bfseries or} & 2 & 4 & 8 & 1 & 9 & 0 & 0 & 0 & 7 & 0 & 0 & 10 & 0 & 0 & 0 & 5 & 0 & 6 & 0 & 3\\
{\bfseries probabl} & 0 & 0 & 0 & 0 & 0 & 0 & 0 & 0 & 0 & 0 & 0 & 0 & 10 & 9 & 8 & 0 & 0 & 0 & 0 & 0\\
{\bfseries s} & 0 & 0 & 0 & 0 & 0 & 0 & 0 & 0 & 0 & 0 & 0 & 0 & 0 & 10 & 9 & 0 & 0 & 0 & 0 & 0\\
{\bfseries that} & 0 & 0 & 0 & 0 & 0 & 0 & 0 & 0 & 0 & 0 & 0 & 0 & 0 & 0 & 10 & 0 & 0 & 0 & 0 & 0\\
{\bfseries they} & 7 & 9 & 8 & 6 & 1 & 7 & 9 & 0 & 0 & 0 & 6 & 2 & 5 & 4 & 3 & 20 & 4 & 0 & 8 & 8\\
{\bfseries well} & 0 & 2 & 6 & 0 & 7 & 0 & 0 & 0 & 5 & 0 & 0 & 8 & 0 & 0 & 0 & 3 & 10 & 4 & 9 & 1\\
{\bfseries were} & 6 & 8 & 0 & 5 & 0 & 0 & 0 & 0 & 0 & 0 & 0 & 0 & 4 & 3 & 2 & 9 & 0 & 10 & 0 & 7\\
{\bfseries women} & 1 & 4 & 12 & 0 & 14 & 0 & 0 & 0 & 10 & 0 & 0 & 16 & 0 & 0 & 0 & 6 & 9 & 8 & 28 & 2\\
{\bfseries year} & 9 & 0 & 0 & 8 & 0 & 0 & 0 & 0 & 0 & 0 & 0 & 0 & 7 & 6 & 5 & 0 & 0 & 0 & 0 &10\\
\end{tabular}}
\end{table}

Each square matrix is calculated moving a window, representing the considered context, over the document, stem by stem. All stems within the window co-occur with the last stem with a strength which is inversely proportional to the distance between the stems. We finally sum the different occurrences of stems: for example, women occurs two times in our document.

\subsection{Cosine similarity}
 In a HAL matrix, rows and columns differ. A word-vector is then represented by the concatenation between the corresponding row and column vectors. In this way we obtain the word-vectors of the keywords we are interested in (white, black, women): $\ket{w_{white}}$, $\ket{w_{black}}$, $\ket{w_{women}}$. We can now calculate the angle between any two word-vectors as well as their cosine similarity ($cs$), since ``cosine has the nice property that it is 1.0 for identical vectors and 0.0 for orthogonal vectors'' \cite{singhal2001modern}. Usually, cosine similarity measures  the similarity between the query vector and the document vector. For this reason, the way we use it, measuring cosine similarity between the keywords in each document, and obtaining each time a different measure, could seem rather unorthodox. However, since the two keywords are just vectors, their angle can be used to measure their similarity in the particular semantic space corresponding to a certain document. For example, as Song, Bruza, and Cole wrote, ``nurse and doctor are similar in semantics to each other, as they always experience the same contexts, i.e., hospital, patients, etc.''\cite{Song}. The reason why we choose to compare cosine similarity to the expectation degree measured through the Born rule is perhaps not intuitive. In both case we deal with many-dimensional vectors, and not with punctiform events. For this reason we will not consider the euclidean distance to calculate similarity or different methods to calculate frequency, such as pointwise mutual information (PPMI).

 \subsection{Gram-Schmidt orthogonalisation} 
In order to measure correlation between two keyword-vectors, let us say \textit{black} and \textit{women}, we first obtain a document vector $\ket{\Psi}$ summing all word-vectors. The next step is to apply the Gram-Schmidt orthogonalisation process to $\ket{w_{black}}$, $\ket{w_{women}}$ in order to obtain two pairs of orthonormal bases $\{\ket{u_{black}}, \ket{u_{black\bot}}\}$ and $\{\ket{u_{women}},\ket{u_{women\bot}}\}$. If we project and normalise the document-vector $\ket{\Psi}$ onto each couple of bases we obtain a vector $\ket{\phi}$:

\begin{equation}
\ket{\phi} = \alpha\ket{u_{black}} + \alpha_\bot\ket{u_{black\bot}} = \beta\ket{u_{women}} + \beta_\bot\ket{u_{women\bot}} 
\end{equation}

We want to emphasize that we represented the document vector through its components on the two bases provided by each keyword-vector.

\subsection{Abstract machines} 
The notion of abstract machine links quantum theory\cite{susskind2014quantum} to post-structuralist perspectives on meaning\cite{deleuze1988thousand}. To typify the semantic relation between \textit{black} and \textit{women} in our corpus, we design two abstract machines: $\sigma$ and $\tau$, two linear operators. Their input vector is $\ket{\phi}$  (representing the document. The $\sigma$-machine operates on each context, and returns the output $+1$ when it acts on the vector $\ket{u_{black}}$ (representing the meaning of the stem \textit{black}),  $-1$ in the other case. In a similar way, the $\tau$-machine applies the same transformation on the meaning of the stem \textit{women}.
Now let us imagine what happens when we apply both the machines to the document: $\sigma\tau\ket{\phi}$. Principally, we deal with three situations:
\begin{enumerate}
	\item the two outcomes are correlated in every context, when the first output is $+1$ and the second is $+1$, and when the first is $-1$, the second is $-1$. If we multiply the two numbers we always score $+1$;
	\item the two outcomes are anti-correlated: in every context where the first output is $+1$, the second will be $-1$ and also the other way round. If we multiply the two numbers, we will always score $-1$;
	\item the two outcomes are not correlated. in some contexts the output of the two machines will be $\{+1, +1\}$, while in others it will be $\{+1, -1\}$, $\{-1, +1\}$, or $\{-1, -1\}$. The average of the outcomes in the different contexts of the considered document will be 0;
\end{enumerate}
The three considered cases are extreme situations: we will also find weak correlations, in which the score will tend to $1$, weak anti-correlation, where the score will tend to $-1$, and also absence of correlation, giving results near $0$. The outcome of a generic machine can be a transformation or not, $+$ or $-$. Since we have two machines, we deal with a four-state semantic space $\sigma\tau = \{++, +-, -+,--\}$. To construct an example of a general machine, we use the following Pauli spin matrix: 

\begin{equation}
\hat{\sigma}_x = \begin{pmatrix}
0 & 1\\ 
1 & 0
\end{pmatrix}
\end{equation}

The effect of this matrix is to switch the components of the state-vector to which it is applied. It is the equivalent of the logical gate \textit{negation} in Quantum Computation\cite{nielsen2004quantum}. In this way we define an operator $\hat{B}_x $ in the \textit{black-base} $\left \{ \ket{u_{black}}, \ket{u_{black\bot}}\right \}$, and an operator $\hat{W}_x $ in the \textit{women-base} $\left \{ \ket{u_{women}}, \ket{u_{women\bot}}\right \}$. As a result, for example, $\hat{W}_x$ switches all the $\ket{u_{women}}$-related components of $\ket{\phi}$:
\begin{equation}
\hat{W}_x(\beta\ket{u_{women}} + \beta_\bot\ket{u_{women\bot}}) = \beta\bot\ket{u_{women}} + \beta\ket{u_{women\bot}}
\end{equation}
$\hat{B}_x$ acts in the same manner on the $\ket{u_{black}}$-related values of $\ket{\phi}$. To calculate the expected score of the application of both machines to the document-vector, we apply the Born rule, whose output will be a number between $-1$ and $1$.

\begin{equation}
r = \bra{\phi}\hat{B}_x\hat{W}_x\ket{\phi} 
\end{equation}

The more the two layers of meaning are connected, the more the two independent abstract machines will return similar outputs: thus we interpret $r$ as the immanent correlation between the respective meanings expressed by the \textit{black} and the \textit{women} stems.

\section{Comparing data}
\begin{figure}
\includegraphics[width=\textwidth]{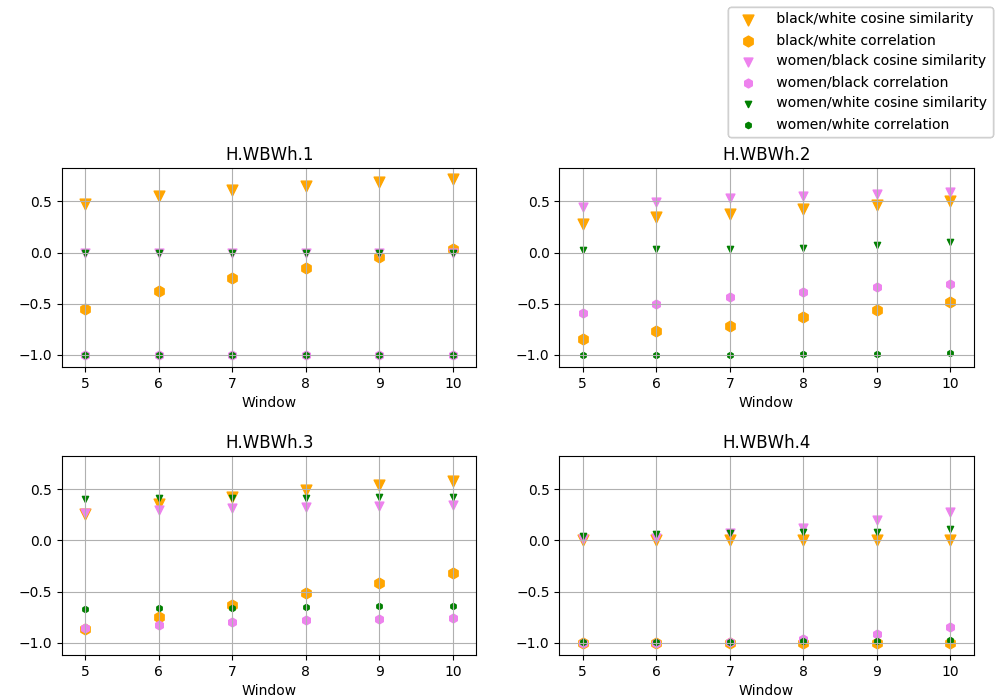}
\caption{Hate speeches H.WBWh.1-4. In H.WBWh.1, \textit{black} and \textit{white} show a high similarity score though they are not correlated. In the text, the meaning of ``white'' is modified both by the prefix \textit{anti-} and by the presence of \textit{black}.}
\label{fig1}
\end{figure}

\begin{figure}
\includegraphics[width=\textwidth]{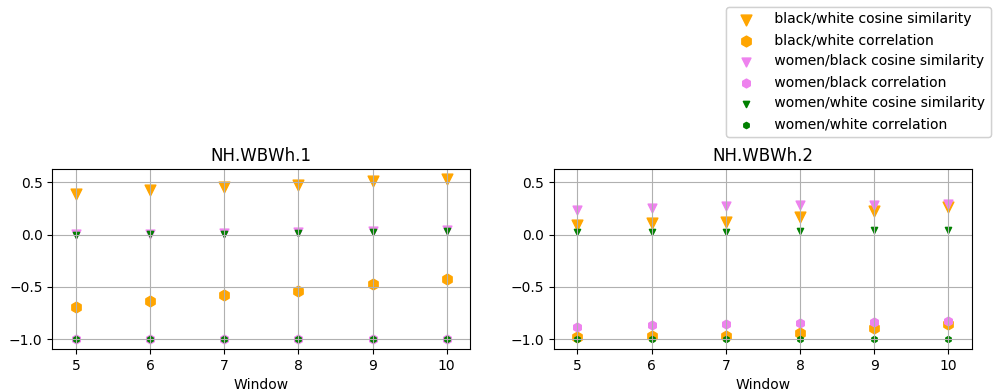}
\caption{Non-hate speeches NH.WBWh.1-2. In NH.WBWh.1, a 0.5 similarity score between \textit{black} and \textit{white} corresponds to a -0.5 value of anticorrelation, since the text is between \textit{black} and \textit{women}. NH.WBWh.1 - a pseudo-scientific argument on IQ, opposes \textit{black} and \textit{white}} \label{fig2}
\end{figure}

\begin{figure}
\centering
\includegraphics[width=\textwidth]{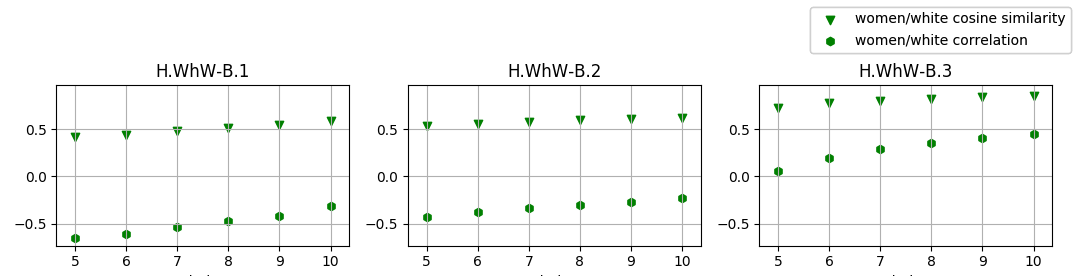}
\caption{Hate speeches H.WhW-B.1-3. In particular, in H.WhW-B.1-2  the keywords occur without a strong relation, whereas H.WhW-B.3 is explicitly on \textit{white women}} \label{fig3}
\end{figure}

\begin{figure}
\includegraphics[width=\textwidth]{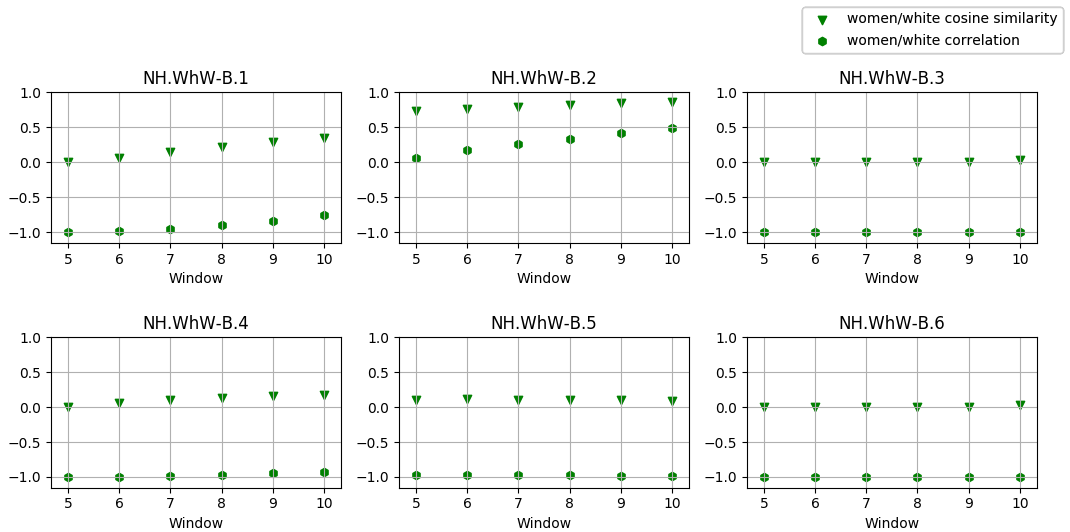}
\caption{Non-hate speeches NH.WhW-B.1-6.  Five out of six documents show a maximal anticorrelation and a 0 similarity score. NH.WhW-B.2 is about \textit{white women} (``Ivanka Voters'').} \label{fig4}
\end{figure}

\begin{figure}
\includegraphics[width=1\textwidth]{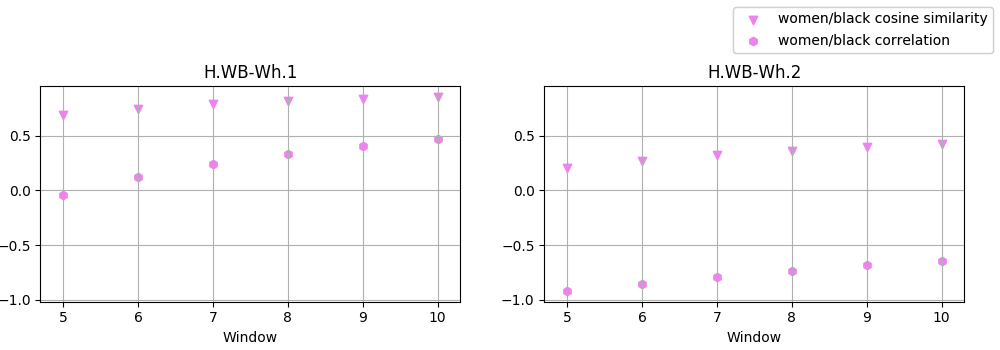}
\caption{Hate speeches H.WB-Wh.1-2. H.WB-Wh.1 is about \textit{black women}, opposed to \textit{black men}; H.WB-Wh.2 carries on an analogy between \textit{blacks' rights} and \textit{women's rights}} \label{fig5}
\end{figure}

\begin{figure}
\includegraphics[width=1\textwidth]{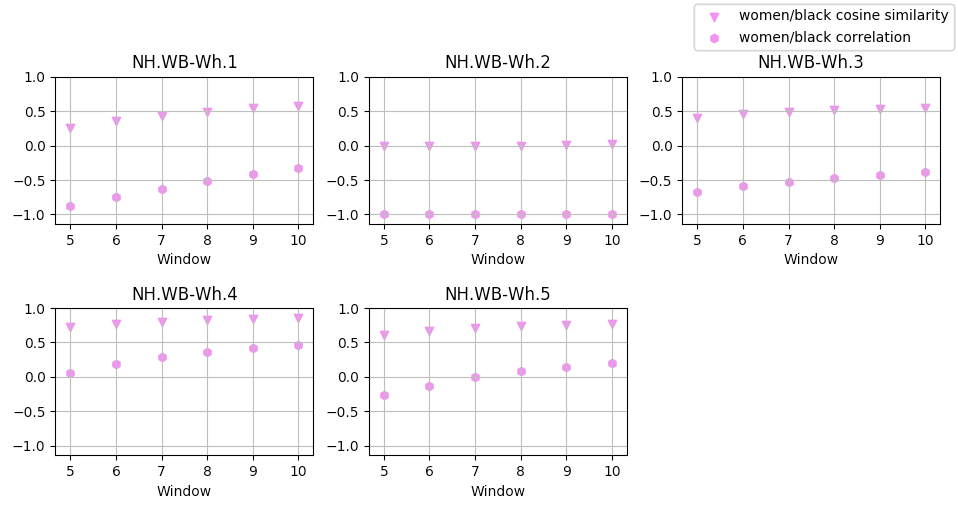}
\caption{Non-hate speeches NH.WB-Wh.1-5 show a great differentiation. In these cases, correlation is helpful to characterise more explicitly what does ``similarity'' mean} \label{fig6}
\end{figure}

We measured cosine similarity and correlation in our corpus of hate and non-hate speeches corresponding to the logical query \textit{black*white*women}. We measured similarity and correlation between \textit{black}, \textit{women} and \textit{white}, \textit{women} respectively. The window length varies from 4 to 10. The results are displayed in fig.~\ref{fig1}, referring to hate speeches, and fig.~\ref{fig2} referring to non-hate speeches. Then we calculated the graphs corresponding to cosine similarity and correlation between \textit{white}, \textit{women} in the corpora of both hate and non-hate speeches where the term \textit{black} is absent - see figures ~\ref{fig3} and ~\ref{fig4}. Finally, we applied the same procedure to \textit{black}, \textit{women} in the corpora of both hate and non-hate speeches where the term \textit{white} is absent - see figures ~\ref{fig5} and ~\ref{fig6}. To compare the results we focus on window lengths of 8-10, which seem associated to more stable values.

\subsection{$black*white*women$}
Both  hate and non-hate speeches present strong anticorrelation \textit{black vs. women} and \textit{white vs. women}. Looking at the document H.BWhW.4, it is not surprising to see that both \textit{black/women} and \textit{white/women} are anticorrelated, since the text draws a comparison between \textit{women's rights} and \textit{black's rights}. We notice also the correspondence between a $ r \simeq 0$ correlation score and a $cs \simeq 0.7$ value of similarity: two word-vectors can be geometrically close without being correlated.
Another interesting problem is the \textit{black/white} opposition. Generally speaking, their anticorrelation is weaker than for the other two, and it tends to disappear in H.WBWh.1 Provided that lexical semantics would describe them as antonyms, it could seem strange that they are not anti-correlated in H.WBWh.1. According to textual semantics \cite{greimas1983structural} semantic relations are modulated and transformed by their co-occurrence in contexts. In our document, \textit{anti-, white,} and \textit{racist} give a fundamental contribution to establish the contextual part of the meaning of \textit{black}, providing it of contextual \textit{classemic values} - Rastier calls them \textit{afferent semes}, and distinguishes them from the \textit{inherent semes}, which characterize the semantic nucleus of a \textit{lexeme}\cite{rastier2009semantique}. In a similar way, H.WBWh.3 is also interesting, since it shows how two terms can be weakly similar ($0.5 \leq cs \leq 0.7$) and still weakly anticorrelated ($ -0.5  \leq r \leq 0$).

\subsection{$white*women-black$}
Five out of six non-hate speeches present a strong anticorrelation \textit{white vs. women}, whereas hate speeches are featured by a weaker anticorrelation; in one case, NH.WhW-B.2, we have a positive correlation, since the document focuses on \textit{white women} voting for D. Trump. If we look at the other documents, they oppose \textit{women} to white men, voters. We must point out how NH.WhW-B.4 could be considered a hate speech from a semantic point of view. In this text, \textit{hispanic women} are opposed to \textit{white girl, men, guys} and this explains the strong anticorrelation.

\subsection{$black*women-white$}
H.WB-Wh.1 presents a positive \textit{correlation} between \textit{black} and \textit{women}: in fact the document opposes black women to black men without reference to white women $(women \to black)$. On the contrary, H.WB-Wh.2 present a strong anticorrelation \textit{black vs. women}, since women and blacks are considered as two distinct minorities. Most non-hate speeches present a weak anticorrelation or a weak correlation, except for NH.WB-Wh.2, in which a maximal anticorrelation value is justified because the document is composed of two different sections, the first about black color and the second about women. In NH.BW-Wh.5 we can see again how a high score of similarity does not necessarily correspond to a correlation of sa given type: in this text, we have a first close co-occurrence of $women$ and $black$; the second occurrence of \textit{women} is free and it weakens the value of the first relation.

\section{Conclusion} 
The paper shows how quantum correlation can clarify the less clear notion of similarity. The terms black and women can occur in hate-speeches with different relations, individuating \textit{black women}, or opposing \textit{women and blacks}. In particular: 
\begin{itemize}
	\item Low similarity values ($0 \leq cs \leq 0.5 $) correspond to a maximum of anticorrelation ($-1 \leq r \leq$ -0.5) between the two stems. We have a double privative semantic opposition $-(A*B)$;
	\item Weak similarity ($0.5 \leq cs \leq 0.7$) correspond to weak anticorrelation or no correlation ($ -0.5  \leq r \leq 0$);
	\item Higher similarity value ($0.7 < cs \leq 1$) corresponds to weak or strong correlations ($0 \leq r \leq 1$): $(A \leftrightarrow B)$;
	
\end{itemize}

The method seems promising as it concerns Digital Humanities and Machine Learning. Many machine learning techniques make use of human beings to label the corpus to avoid to define the involved labels, at the risk of mistakes and ambiguity. Quantum semantics offers a different perspective on meaning, which can be useful to re-classify the corpus. For example, many hate speeches present strong anti-correlations between terms no matter of the width of the window. Furthermore, similar \textit{semantic profiles} such as NH.WhW-B.2, H.WhW-B.3, NH.BW-Wh.4, H.BW-Wh.1 reveal a similar topic (black or white women) and show a sexist connotation, no matter how they have been labelled.

\subsection{Semantics and information}
As we wrote, the method we described applies to Semantics only in a narrow sense (see paragraph ~\ref{sec:relevance}). Actually, the algorithm does not understand meaning: its task is not to translate texts; the algorithm provides information \textit{on} meaning: on textual semantic structure, on some functions of the semantic system that produces the text (in our case: $a \vee b$; $c \rightarrow d$) in the light of the semiotic square \cite{square}. For example, the algorithm could be applied to a document written in a previously unknown language such as Minoan linear A script, or to an encrypted one. As long as we can distinguish its words, the algorithm would not decode the document, but it would provide information on what lexemes can be considered similar, implied or opposed as meaning is regarded, mining information from their co-occurence in the text and measuring the expectations related to some simple transformations operated on the coherent distributions of meaning - \textit{isotopies} \cite{greimas1983structural} - along the text.

\subsection{Future work}
In this paper we measured the correlation with reference to the single Pauli operator $\sigma_x$. To improve the method, we will measure the expectations of Pauli's operators $\sigma_y$ and $\sigma_z$ to get an alternative way to measure entanglement\cite{susskind2014quantum}, to be compared to Bell inequalities used by Barros et al. \cite{bell}. On a similar line, the Born rule allows us to work on density matrices. Thus we hope to get further insights on the relation between Von Neumann information and meaning.

\bibliographystyle{eptcs}
\bibliography{references}

\begin{thebibliography}{10}
\providecommand{\bibitemdeclare}[2]{}
\providecommand{\surnamestart}{}
\providecommand{\surnameend}{}
\providecommand{\urlprefix}{Available at }
\providecommand{\url}[1]{\texttt{#1}}
\providecommand{\href}[2]{\texttt{#2}}
\providecommand{\urlalt}[2]{\href{#1}{#2}}
\providecommand{\doi}[1]{doi:\urlalt{http://dx.doi.org/#1}{#1}}
\providecommand{\bibinfo}[2]{#2}

\bibitemdeclare{}{OHI}
\bibitem{OHI}
 (\bibinfo{year}{2018}): \emph{\bibinfo{title}{The Online Hate Index}}.
\newblock
  \urlprefix\url{http://www.adl.org/resources/reports/the-online-hate-index}.

\bibitemdeclare{inproceedings}{bell}
\bibitem{bell}
\bibinfo{author}{Jo{\~a}o \surnamestart Barros\surnameend},
  \bibinfo{author}{Zeno \surnamestart Toffano\surnameend},
  \bibinfo{author}{Youssef \surnamestart Meguebli\surnameend} \&
  \bibinfo{author}{Bich-Li{\^e}n \surnamestart Doan\surnameend}
  (\bibinfo{year}{2014}): \emph{\bibinfo{title}{Contextual Query Using Bell
  Tests}}.
\newblock In \bibinfo{editor}{Harald \surnamestart Atmanspacher\surnameend},
  \bibinfo{editor}{Emmanuel \surnamestart Haven\surnameend},
  \bibinfo{editor}{Kirsty \surnamestart Kitto\surnameend} \&
  \bibinfo{editor}{Derek \surnamestart Raine\surnameend}, editors: {\sl
  \bibinfo{booktitle}{Quantum Interaction}}, \bibinfo{publisher}{Springer
  Berlin Heidelberg}, \bibinfo{address}{Berlin, Heidelberg}, pp.
  \bibinfo{pages}{110--121}, \doi{10.1007/978-3-642-54943-4_10}.

\bibitemdeclare{conference}{Bruza}
\bibitem{Bruza}
\bibinfo{author}{Peter \surnamestart Bruza\surnameend} \& \bibinfo{author}{John
  \surnamestart Woods\surnameend} (\bibinfo{year}{2008}):
  \emph{\bibinfo{title}{Quantum collapse in semantic space : interpreting
  natural language argumentation}}.
\newblock In: {\sl \bibinfo{booktitle}{Proceedings of the Second Quantum
  Interaction Symposium (QI-2008)}}, \bibinfo{publisher}{College Publications},
  pp. \bibinfo{pages}{141--147}.

\bibitemdeclare{article}{clauser1969proposed}
\bibitem{clauser1969proposed}
\bibinfo{author}{John~F \surnamestart Clauser\surnameend},
  \bibinfo{author}{Michael~A \surnamestart Horne\surnameend},
  \bibinfo{author}{Abner \surnamestart Shimony\surnameend} \&
  \bibinfo{author}{Richard~A \surnamestart Holt\surnameend}
  (\bibinfo{year}{1969}): \emph{\bibinfo{title}{Proposed experiment to test
  local hidden-variable theories}}.
\newblock {\sl \bibinfo{journal}{Physical review letters}}
  \bibinfo{volume}{23}(\bibinfo{number}{15}), p. \bibinfo{pages}{880},
  \doi{10.1103/PhysRevLett.23.880}.

\bibitemdeclare{book}{deleuze1988thousand}
\bibitem{deleuze1988thousand}
\bibinfo{author}{Gilles \surnamestart Deleuze\surnameend} \&
  \bibinfo{author}{F{\'e}lix \surnamestart Guattari\surnameend}
  (\bibinfo{year}{1988}): \emph{\bibinfo{title}{A thousand plateaus: Capitalism
  and schizophrenia}}.
\newblock \bibinfo{publisher}{Bloomsbury Publishing}.

\bibitemdeclare{inproceedings}{eigen}
\bibitem{eigen}
\bibinfo{author}{Fran{\c{c}}ois \surnamestart Dubois\surnameend} \&
  \bibinfo{author}{Zeno \surnamestart Toffano\surnameend}
  (\bibinfo{year}{2017}): \emph{\bibinfo{title}{Eigenlogic: A Quantum View for
  Multiple-Valued and Fuzzy Systems}}.
\newblock In \bibinfo{editor}{Jose~Acacio \surnamestart de~Barros\surnameend},
  \bibinfo{editor}{Bob \surnamestart Coecke\surnameend} \&
  \bibinfo{editor}{Emmanuel \surnamestart Pothos\surnameend}, editors: {\sl
  \bibinfo{booktitle}{Quantum Interaction}}, \bibinfo{publisher}{Springer
  International Publishing}, \bibinfo{address}{Cham}, pp.
  \bibinfo{pages}{239--251}, \doi{10.1007/978-3-642-54943-4_10}.

\bibitemdeclare{book}{Eco}
\bibitem{Eco}
\bibinfo{author}{Umberto \surnamestart Eco\surnameend} (\bibinfo{year}{2014}):
  \emph{\bibinfo{title}{From the tree to the labyrinth}}.
\newblock \bibinfo{publisher}{Harvard University Press},
  \doi{10.4159/9780674728165}.

\bibitemdeclare{article}{doi:10.1108/K-05-2017-0187}
\bibitem{doi:10.1108/K-05-2017-0187}
\bibinfo{author}{Francesco \surnamestart Galofaro\surnameend},
  \bibinfo{author}{Zeno \surnamestart Toffano\surnameend} \&
  \bibinfo{author}{Bich-Li{\^e}n \surnamestart Doan\surnameend}
  (\bibinfo{year}{2018}): \emph{\bibinfo{title}{A quantum-based semiotic model
  for textual semantics}}.
\newblock {\sl \bibinfo{journal}{Kybernetes}}
  \bibinfo{volume}{47}(\bibinfo{number}{2}), pp. \bibinfo{pages}{307--320},
  \doi{10.1108/K-05-2017-0187}.

\bibitemdeclare{article}{square}
\bibitem{square}
\bibinfo{author}{A.~J. \surnamestart Greimas\surnameend} \&
  \bibinfo{author}{Fran{\c c}ois \surnamestart Rastier\surnameend}
  (\bibinfo{year}{1968}): \emph{\bibinfo{title}{The Interaction of Semiotic
  Constraints}}.
\newblock {\sl \bibinfo{journal}{Yale French Studies}} (\bibinfo{number}{41}),
  pp. \bibinfo{pages}{86--105}, \doi{10.2307/2929667}.
\newblock \urlprefix\url{http://www.jstor.org/stable/2929667}.

\bibitemdeclare{book}{greimas1983structural}
\bibitem{greimas1983structural}
\bibinfo{author}{Algirdas~Julien \surnamestart Greimas\surnameend}
  (\bibinfo{year}{1983}): \emph{\bibinfo{title}{Structural semantics: An
  attempt at a method}}.
\newblock \bibinfo{publisher}{University of Nebraska Press}.

\bibitemdeclare{article}{Lund1996}
\bibitem{Lund1996}
\bibinfo{author}{Kevin \surnamestart Lund\surnameend} \& \bibinfo{author}{Curt
  \surnamestart Burgess\surnameend} (\bibinfo{year}{1996}):
  \emph{\bibinfo{title}{Producing high-dimensional semantic spaces from lexical
  co-occurrence}}.
\newblock {\sl \bibinfo{journal}{Behavior Research Methods, Instruments, {\&}
  Computers}} \bibinfo{volume}{28}(\bibinfo{number}{2}), pp.
  \bibinfo{pages}{203--208}, \doi{10.3758/BF03204766}.

\bibitemdeclare{book}{Melucci}
\bibitem{Melucci}
\bibinfo{author}{Massimo \surnamestart Melucci\surnameend}
  (\bibinfo{year}{2015}): \emph{\bibinfo{title}{Introduction to information
  retrieval and quantum mechanics}}.
\newblock \bibinfo{publisher}{Springer}, \bibinfo{address}{Berlin, Heidelberg},
  \doi{10.1007/978-3-662-48313-8}.

\bibitemdeclare{book}{nielsen2004quantum}
\bibitem{nielsen2004quantum}
\bibinfo{author}{Michael~A \surnamestart Nielsen\surnameend} \&
  \bibinfo{author}{Isaac~L \surnamestart Chuang\surnameend}
  (\bibinfo{year}{2004}): \emph{\bibinfo{title}{Quantum Computation and Quantum
  Information (Cambridge Series on Information and the Natural Sciences)}}.
\newblock \bibinfo{publisher}{Cambridge university press}.

\bibitemdeclare{book}{petitot2004morphogenesis}
\bibitem{petitot2004morphogenesis}
\bibinfo{author}{Jean \surnamestart Petitot\surnameend} (\bibinfo{year}{2004}):
  \emph{\bibinfo{title}{Morphogenesis of meaning}}.
\newblock \bibinfo{publisher}{P. Lang}.

\bibitemdeclare{book}{rastier2009semantique}
\bibitem{rastier2009semantique}
\bibinfo{author}{Fran{\c{c}}ois \surnamestart Rastier\surnameend}
  (\bibinfo{year}{2009}): \emph{\bibinfo{title}{S{\'e}mantique
  interpr{\'e}tative}}.
\newblock \bibinfo{publisher}{Presses universitaires de France},
  \doi{10.3917/puf.rast.2009.01}.

\bibitemdeclare{article}{singhal2001modern}
\bibitem{singhal2001modern}
\bibinfo{author}{Amit \surnamestart Singhal\surnameend} (\bibinfo{year}{2001}):
  \emph{\bibinfo{title}{Modern Information Retrieval: {A} Brief Overview}}.
\newblock {\sl \bibinfo{journal}{{IEEE} Data Eng. Bull.}}
  \bibinfo{volume}{24}(\bibinfo{number}{4}), pp. \bibinfo{pages}{35--43}.
\newblock \urlprefix\url{http://sites.computer.org/debull/A01DEC-CD.pdf}.

\bibitemdeclare{article}{Song}
\bibitem{Song}
\bibinfo{author}{Dawei \surnamestart Song\surnameend}, \bibinfo{author}{Dawei
  \surnamestart Song\surnameend}, \bibinfo{author}{Peter \surnamestart
  Bruza\surnameend} \& \bibinfo{author}{Richard \surnamestart Cole\surnameend}
  (\bibinfo{year}{2004}): \emph{\bibinfo{title}{Concept Learning and
  Information Inferencing on a High Dimensional Semantic Space}}.
\newblock {\sl \bibinfo{journal}{In: ACM SIGIR 2004 Workshop on
  Mathematical/Formal Methods in Information Retrieval (MF/IR2004). Sheffield,
  United Kingdom, 25-29 July 2004.}}, \doi{10.1.1.370.4676}.

\bibitemdeclare{book}{susskind2014quantum}
\bibitem{susskind2014quantum}
\bibinfo{author}{Leonard \surnamestart Susskind\surnameend} \&
  \bibinfo{author}{Art \surnamestart Friedman\surnameend}
  (\bibinfo{year}{2014}): \emph{\bibinfo{title}{Quantum mechanics: the
  theoretical minimum}}.
\newblock \bibinfo{publisher}{Basic Books (AZ)}.

\bibitemdeclare{book}{van2004geometry}
\bibitem{van2004geometry}
\bibinfo{author}{Cornelis~Joost \surnamestart Van~Rijsbergen\surnameend}
  (\bibinfo{year}{2004}): \emph{\bibinfo{title}{The geometry of information
  retrieval}}.
\newblock \bibinfo{publisher}{Cambridge University Press},
  \doi{10.1017/CBO9780511543333}.

\bibitemdeclare{book}{Zinoviev}
\bibitem{Zinoviev}
\bibinfo{author}{Dmitry \surnamestart Zinoviev\surnameend}
  (\bibinfo{year}{2016}): \emph{\bibinfo{title}{Data Science Essentials in
  Python: Collect-Organize-Explore-Predict-Value}}.
\newblock \bibinfo{publisher}{Pragmatic Bookshelf}.

\end{thebibliography}
\end{document}